\documentstyle[11pt]{article}
\newlength{\mytopmargin}
\newlength{\myleftmargin}
\begin{document}
\title{
Global fluctuation formulas and universal correlations
for random matrices and log-gas systems at infinite density}
\author{P.J. Forrester\thanks{Supported by the Australian Research Council}
\\
Department of Mathematics \\
University of Melbourne \\
Parkville, Victoria  3052 \\
Australia}

\date{}
\maketitle
\begin{abstract} It is shown how the universal correlation function of
Br\'ezin and Zee, and Beenakker, for random matrix ensembles of Wigner-Dyson
type with density support on a finite interval can be derived using a linear
response argument and macroscopic electrostatics. The analogous formula for
ensembles of unitary random matrices is derived using the same method, and
the corresponding global fluctuation formula for the variance of a linear
statistic is presented. The result for the universal correlation is checked
using
an exact result for the finite system two-point correlation
in the Dyson circular ensemble
at all even $\beta$.
\\
{}
\end{abstract}

\section{Introduction}
\setcounter{equation}{0}
\renewcommand{\theequation}{1.\arabic{equation}}

A large portion of random matrix theory is concerned with the eigenvalue
probability density function
\begin{equation}
\prod_{l=1}^Ne^{-\beta V(x_l)} \prod_{1 \le j < k \le N} |x_k - x_j|^\beta,
\qquad x_j \in
{\bf R}
\end{equation}
where $\beta = 1, 2$ or 4 according to the matrices being real symmetric,
Hermitian or quaternion real respectively (see e.g. [1]). An alternative
interpretation of (1.1) is as the Boltzmann factor of a one-component
log-potential Coulomb system confined to a line with potential energy
\begin{equation}
\sum_{l=1}^N V(x_l) - \sum_{1 \le j < k \le N} \log |x_k - x_j|
\end{equation}
The two body term in (1.2) is the electrostatic energy of the Coulomb repulsion
between like charges, while the one body potential acts to confine the
particles to some interval of the real line.

In particular, consider the situation in which $V(x)$ is infinite outside an
interval $(a,b)$ so that
\begin{equation}
x_j \in (a,b)
\end{equation}
This setting has recently been discussed by Br\'{e}zin and Zee [2], in the
special
case $\beta = 2$ and $V(x)$ an even polynomial about the centre of the interval
(see also [3]),
and by Beenakker [4], for general $\beta$ and $V(x)$. These authors have found
that the two-point correlation, $\rho^T(x,x')$ say, satisfies the universal
formula
$$
\lim_{N \rightarrow \infty} {N \over b-a}
\int_{x'}^{x'+(b-a)/N} dt \, \rho_N^T(x,t) \hspace{8cm}
$$
\begin{equation}
=-{1 \over \beta \pi^2 (x - x')^2} {(a+b)(x+x')/2 - ab - xx' \over
[(x - a)(b - x)(x' - a)(b - x')]^{1/2}}, \quad x \not= x'
\end{equation}

\noindent
Remark: The function $\rho^T(x,x')$ above equals $-T_2(x,x')$ in the notation
of Beenakker [4].\\
The l.h.s. of (1.4) corresponds to the two point correlation in the infinite
density limit, smoothed over the average interparticle spacing.

As a consequence of (1.4), Beenakker [4] has derived the global fluctuation
formula
\begin{eqnarray}
\lefteqn{\lim_{N \rightarrow \infty} {\rm var} \left [ \sum_{j=1}^N f(x_j)
 \right ]}\nonumber \\
& & = {1 \over \beta \pi^2 } {\cal P} \int_a^b dx \int_a^b dx'
\left [ {(x' - a)(b - x') \over (x - a)(b - x)} \right ]^{1/2} {f(x) \over
x - x'} {d \over dx'} f(x')
\end{eqnarray}
where ${\cal P}$ denotes the principal part,
for the variance of a linear statistic.

The study of random matrix ensembles with density support confined to a finite
interval follows the study [5-9] of ensembles with density support on the half
interval $[0, \infty)$, which are of relevance to fluctuation phenomena in
mesoscopic systems. In this setting it is not necessary  to
consider an infinite density limit. Rather, a scale can be chosen so that the
number of eigenvalues near the``edge" $x=0$ remains finite [7] in the
$N \rightarrow \infty$ limit. The asymptotic expansion of the corresponding
two-point correlation satisfies a formula analogous to (1.3):
\begin{equation}
\rho^T(x,x') \: \sim \: - {1 \over 2 \pi^2 \beta \sqrt{x x'}} {x + x' \over
(x - x')^2}
\end{equation}
for the leading non-oscillatory behaviour. Beenakker [9] (see also [10])
has used (1.5) to derive the analogue of (1.4) for the fluctuation of a
linear statistic.

Recently [11] a new derivation of (1.5) has been given, applicable to all
log-potential Coulomb systems in their conductive phase, which is based on
macroscopic electrostatics and linear response theory. It is our purpose
herein to first rederive (1.4) using the method of [11], and then to consider
the analogues of (1.4) and (1.5) for unitary random matrices (or equivalently
log-gas systems confined to a circle). For the one-component log-gas on a
circle at even $\beta$, the two point correlation has been evaluated exactly
[12] in terms of a
$\beta$-dimensional integral for all particle numbers $N$, which allows an
explicit verification of the
analogue of (1.4).

\section{Macroscopic electrostatic argument}
\setcounter{equation}{0}
\renewcommand{\theequation}{2.\arabic{equation}}

Our objective in this section is to derive (1.4). For simplicity of notation we
consider the symmetric case $a=-b$ and scale the variables $x \mapsto bx$ and
$x' \mapsto bx'$, so that the spectrum has support in ($-$1,1). The r.h.s. of
(1.4) then reads
\begin{equation}
-{1 \over \beta \pi^2 (x - x')^2 } {1 - x x' \over [(1 - x^2) (1 - x'^2)]^{1
\over
2}}, \qquad x \not= x'
\end{equation}
There is no loss of generality is specializing to this case as the r.h.s. of
(1.4) can be reclaimed from (2.1) by rescaling the variables
$x \mapsto 2(x - (a+b)/2)/(b-a)$ (and similarly $x'$).
Rather than restricting ourselves to systems with probability density function
(1.1) we will consider any log-potential Coulomb system confined to the
interval
($-$1,1) which consists of point particles and is in a conductive phase.

Our starting point is the linear response argument of
 [11], which says that if the system is perturbed by adding an external charge
$\delta q$ at a point $\vec{r'}$, then
\begin{equation}
<\Phi(\vec{r})> -<\Phi(\vec{r})>_0 =
-\beta \delta q <\Phi(\vec{r})\Phi(\vec{r'}) >^T
\end{equation}
where $\Phi(\vec{r})$ is the potential at the point $\vec{r}$ due to the
induced charge distribution, $< >_0$ refers to the value of the average before
the perturbation and $<AB>^T=<AB>_0-<A>_0<B>_0$.

To obtain from (2.2) the charge density- charge density correlation
\begin{equation}
<\sigma(x)\sigma(x')>^T
\end{equation}
we hypothesize that in the high density $N \rightarrow \infty$ limit, the
system
behaves as a conducting interval from $-$1 to 1 along the $x$-axis, obeying
the laws of two-dimensional macroscopic electrostatics. Now, for a point
$x$ on the $x$-axis, macroscopic electrostatics  gives
\begin{equation}
E^+_y(x)-E^-_y(x)=2 \pi \sigma (x),
\end{equation}
where the $y$-components of the electric fields are given in terms of the
potential by
\begin{equation}
E^+_y(x) := {\partial \over \partial y} \Phi (\vec{r}) \Big |_{y = 0^+}
\quad
E^-_y(x) := {\partial \over \partial y} \Phi (\vec{r}) \Big |_{y = 0^-}
\end{equation}
Hence
\begin{eqnarray}
\lefteqn{-\beta \delta q <\sigma(x)\sigma(x')>^T}\nonumber \\
& = & - \beta \delta q {2 \over (2 \pi)^2}
\left ( <E^+_y(x)E^+_{y'}(x')>^T - <E^+_y(x)E^-_{y'}(x')>^T \right )
\nonumber\\
& = & {2 \over (2 \pi)^2} \left (
{\partial \over \partial y}{\partial \over \partial y'} \big (
<\Phi(\vec{r})> -<\Phi(\vec{r})>_0 \big) \Big |_{y=0^+,y'=0^+} \right
.\nonumber\\
& & - \left .{\partial \over \partial y}{\partial \over \partial y'} \big (
<\Phi(\vec{r})> -<\Phi(\vec{r})>_0 \big) \Big |_{y=0^+,y'=0^-} \right )
\end{eqnarray}
where to obtain the first line we have used the symmetry
$$
<E^+_y(x)E^s_{y'}(x')>^T = <E^-_y(x)E^{-s}_{y'}(x')>^T
$$
for $s = +,-$ and to obtain the second line (2.5) and (2.2) have been used.

By the hypothesis of the applicability of macroscopic electrostatics,
$$
<\Phi(\vec{r})> -<\Phi(\vec{r})>_0
$$
is simply the potential at a point $\vec{r}$ due to the charge induced on the
conducting interval \\
($-$1,1) by the external charge $\delta q$. The calculation
of this potential is a standard problem in $2d$ electrostatics [13,
Chapter 4, exercise 5]. One first
considers the potential at $\vec{r}$, $|\vec{r}|>1$, due to a unit charge at
$\vec{r'}$,
$|\vec{r'}|>1$, and a conducting unit disk centred at the origin. By the
method of images, this potential is
\begin{equation}
\Phi (z,z') = - \log \left ( |z' - z| /|z' - 1/z^*| \right )-\log |z'|
\end{equation}
where $z\, (z')$ is the complex coordinate corresponding to $\vec{r} = (x,y)
\, (\vec{r'} = (x',y'))$, and it is assumed the conductor carries no net
charge. Then one maps the exterior of the unit disk to the
plane, with a cut from $-$1 to 1 along the $x$-axis by the conformal mapping
\begin{equation}
w = z +{1 \over z} \quad {\rm and \:\: thus}  \quad z = w + (w^2 - 1)^{1/2},
\end{equation}
where the principal branch of the square root is chosen.

Using (2.8) to substitute in (2.7) for $z$ (and similarly $z'$) we obtain the
potential
at $\vec{r}$ due to the charge induced on the conducting interval and due to
the
external charge at $\vec{r'}$. Subtracting the  potential
due to the external charge we therefore
have
\begin{eqnarray}
\lefteqn{<\Phi(\vec{r})> -<\Phi(\vec{r})>_0} \nonumber \\
& = & \delta q \Big ( \log|w - w'| - \log|w'-w+(w'^2-1)^{1/2}-(w^2-1)^{1/2}|
\nonumber
\end{eqnarray}
\begin{equation}
+ \log|w'-w^*+(w'^2-1)^{1/2}+((w^2-1)^{1/2})^*|-\log |w'+(w'^2-1)^{1/2}| \big )
\end{equation}
where
\begin{equation}
w = x + iy \qquad {\rm and} \qquad w' = x' + i y'
\end{equation}
and we have used the formula
\begin{equation}
1/ (w + (w^2 - 1)^{1/2})  =  w - (w^2 - 1)^{1/2}
\end{equation}

Substituting (2.9) in (2.6), it remains to perform the derivatives. Now, for
$-1<x<1$ and $0<y\ll1$
\begin{equation}
  (w^2 - 1)^{1/2} \: \sim \:  |x^2 - 1|^{1/2}i e^{i x y /(x^2-1)}
\end{equation}
so differentiating (2.9) with respect to $y$ gives
\begin{eqnarray}
\lefteqn{{\partial \over \partial y}
\big (
<\Phi(\vec{r})> -<\Phi(\vec{r})>_0 \big) \Big |_{y=0^+}}\nonumber \\
&= &\delta q \left ( -{\rm Re} {i \over w' - x} - {2\over |1 - x^2|^{1/2}}
{\rm Re} \left [ {u \over w' + (w'^2 - 1)^{1/2} - u} \right ] \right )
\end{eqnarray}
where
\begin{equation}
u := x + i|x^2 - 1|^{1/2}
\end{equation}
Using (2.11) with $w,x,y$ replaced by $w',x',y'$, (2.12) then gives
\begin{eqnarray}
\lefteqn{{\partial \over \partial y}{\partial \over \partial y'} \big (
<\Phi(\vec{r})> -<\Phi(\vec{r})>_0 \big) \Big |_{y=0^+,y'=0^+} } \nonumber \\
& & = \delta q \left ( - {1 \over (x' - x)^2} + {2\over |1 - x^2|^{1/2}|1 -
x'^2|^{1/2}}
{\rm Re} \left [ {u'u \over (u'  - u)^2} \right ] \right )
\end{eqnarray}
where
\begin{equation}
u' := x' + i|x'^2 - 1|^{1/2}
\end{equation}
Also, since for $-1 < x' < 1$ and $-1 \ll y < 0$
\begin{equation}
 (w'^2 - 1)^{1/2} \: \sim \:
- |x'^2 - 1|^{1/2}i e^{i x' y' /(x'^2-1)},
\end{equation}
we  have
\begin{eqnarray}
\lefteqn{{\partial \over \partial y}{\partial \over \partial y'} \big (
<\Phi(\vec{r})> -<\Phi(\vec{r})>_0 \big) \Big |_{y=0^+,y'=0^-} } \nonumber \\
& & = \delta q \left ( - {1 \over (x' - x)^2} + {2\over |1 - x^2|^{1/2}|1 -
x'^2|^{1/2}}
{\rm Re} \left [ {u'^*u \over (u'^*  - u)^2} \right ] \right )
\end{eqnarray}

We now substitute the evaluated derivatives (2.15) and (2.18) in (2.6) to
obtain for
the charge-charge correlation
\begin{eqnarray}
\lefteqn{<\sigma(x)\sigma(x')>^T} \nonumber \\
& = & {1 \over \beta \pi^2 |1 - x^2|^{1/2}|1 - x'^2|^{1/2}}
{\rm Re} \left [{u'u \over (u'  - u)^2} +
{u'^*u \over (u'^*  - u)^2} \right ]
\end{eqnarray}
Straightforward simplification of the r.h.s. of (2.19) using the definitions
(2.14) and (2.16) of $u$ and $u'$ reclaims (2.1).

\section { Ensembles of unitary random matrices}
\setcounter{equation}{0}
\renewcommand{\theequation}{3.\arabic{equation}}

The Dyson circular ensembles of unitary random matrices (see e.g. [1])
have eigenvalue probability density function
\begin{equation}
\prod_{1 \le j < k \le N} |e^{i \theta_k} - e^{i \theta_j}|^{\beta}
\end{equation}
where $\beta =1,2$ or 4 according to the ensemble being invariant under
orthogonal, unitary or symplectic similarity transformations respectively.
The probability density is identical, up to a multiplicative constant, to the
Boltzmann factor of a one-component log-potential Coulomb system confined to
a unit circle. It is particularly simple to apply the argument of Section 2 to
this setting and thus deduce that the corresponding two-point correlation
$\rho^T(\theta,\theta')$ satisfies the universal formula
\begin{equation}
\lim_{N \rightarrow \infty} {N \over 2 \pi} \int_\theta^{\theta' + 2 \pi /N}
d \phi \, \rho^T_N(\theta, \phi)
= -{ 1 \over  \beta (2\pi)^2 \sin^2\Big ((\theta - \theta')/2 \Big )}
\end{equation}
Let us briefly indicate the required working.

First, since the domain is a circle, in polar coordinates
the perpendicular component of the
electric field is the $r$-component, and thus the analogue of (2.4) is
\begin{equation}
E_r^+(\theta) - E_r^-(\theta) = 2 \pi \sigma (\theta)
\end{equation}
Let us now assume that the source point $\vec{r'}$ is outside the unit
circle. Then, if the field point $\vec{r}$ is also outside the circle, the
total
potential created is the same as that with the  circle replaced by a unit
conducting disk and is thus given by (2.7). Subtracting the charge-charge
portion of this potential
then gives
\begin{equation}
<\Phi(\vec{r})> -<\Phi(\vec{r})>_0 = \delta q \Big (\log |z' - 1/z^*|
- \log |z'| \Big )
\end{equation}
For the field point inside the unit circle, there is no net potential due to
the source point outside the circle. Subtracting the charge-charge potential
gives in this case
\begin{equation}
<\Phi(\vec{r})> -<\Phi(\vec{r})>_0 = \delta  q \log |z' - z|
\end{equation}
The analogue of (2.6) can now easily be evaluated  to
derive the r.h.s. of (3.2).

The result (3.2) can be used to derive for
 random unitary matrices (or equivalently, log-potential
Coulomb systems on a circle) the analogue of the global fluctuation
formula (2.6) . For general systems of point particles on a unit
circle
\begin{eqnarray}
\lefteqn{\lim_{N \rightarrow \infty} {\rm var} \left [ \sum_{j=1}^N f(x_j)
 \right ]}\nonumber \\
&= &\lim_{N \rightarrow \infty}
\int_0^{2 \pi} d\theta \int_0^{2 \pi} d\theta' \,f(\theta)f(\theta')
\Big (\rho^T_N(\theta - \theta') + \rho \delta (\sin (\theta - \theta')/2)
\Big )
\end{eqnarray}
where $\rho := N / 2 \pi$ and the delta denotes the Dirac delta function.
Introducing the Fourier series
\begin{equation}
\lim_{N \rightarrow \infty} \left (
\rho^T_N(\theta - \theta') + \rho \delta (\sin (\theta - \theta')/2) \right )
= \sum_{n = -\infty}^\infty s_n e^{i (\theta - \theta')n}
\end{equation}
this reads
\begin{equation}
\lim_{N \rightarrow \infty} {\rm var} \left [ \sum_{j=1}^N f(\theta_j)
 \right ] = \sum_{n = -\infty}^\infty s_n |f_n|^2
\end{equation}
where
\begin{equation}
f_n := \int_0^{2 \pi} d\theta \, f(\theta)e^{i n \theta}
\end{equation}

The coefficients $s_n$ are given by the usual inversion formula of (3.7).
In the inversion formula we replace $\rho^T(\theta)$ by its smoothed
value (3.2) written as a second derivative:
\begin{equation}
\rho^T(\theta) = {1 \over \pi^2 \beta} {d^2 \over d \theta^2}
\log |\sin \theta /2|, \qquad \theta \ne 0
\end{equation}
We then calculate the coefficients by a formal differentiation by parts:
\begin{eqnarray}
s_n & = & {1 \over \pi^2 \beta} {1 \over 2 \pi} \int_{-\pi}
^{\pi} d\theta \, \Big ({d^2 \over d \theta^2}\log |\sin \theta /2|
\Big )e^{i n \theta}
\nonumber \\
& = & {1 \over \pi^2 \beta} (-n^2){1 \over 2 \pi}  \int_{-\pi}
^{\pi} d\theta \, \left (\log |\sin \theta /2| \right )e^{i n \theta}\nonumber
\\
& = & {|n| \over 2\pi^2 \beta}
\end{eqnarray}
Hence, after substituting (3.11) in (3.8), we obtain the desired
global fluctuation
formula
\begin{equation}
\lim_{N \rightarrow \infty} {\rm var} \left [ \sum_{j=1}^N f(\theta_j)
 \right ] = {1 \over 2\pi^2 \beta}\sum_{n = -\infty}^\infty |n| |f_n|^2
\end{equation}

A recent exact calculation  [14] allows the formula (3.2) to be
illustrated for the one-component log-gas on a unit circle at all even values
of $\beta$. The exact calculation gives that the two particle distribution
function
$\rho_N (0,\theta)$ for this system is equal to a $\beta$-dimensional integral
[14, eq. (3.15)\footnote{This equation is missing a factor of $e^{-\pi i\eta
\gamma (N-2)/N}$; c.f. eq. (3.2) of [14].} with $\gamma = \beta$]:
$$
\rho_N(0,\theta) = {N(N-1)\over L^2} (2 \sin \pi \theta/L )^\beta
B_N(\beta) e^{- \pi i \beta \theta (N-2)/L}
$$
\begin{equation}
\times \int_{[0,1]^\beta}dt_1 \dots dt_\beta \,
\prod_{j=1}^\beta[1 - (1 - e^{2 \pi i \theta /L})t_j]^{N-2}
D_{2 /\beta,2 /\beta,4 /\beta}(t_1,\dots,t_\beta)
\end{equation}
where
\begin{equation}
D_{2 /\beta,2 /\beta,4 /\beta}(t_1,\dots,t_\beta) :=
\prod_{j=1}^\beta[(1-t_j)t_j]^{1 - 2/\beta} \prod_{1 \le j < k \le \beta}
|t_k - t_j|^{4 / \beta}
\end{equation}
and $L = 2 \pi$.
The exact value of the normalization $B_N(\beta)$, which is given explicitly in
[14], is not required for our discussion below.

We seek to expand (3.13) for large-$N$ so that the integral on the l.h.s
 of (3.2)
can be evaluated. In the large-$N$ limit we see that the $N$-dependent factors
of the integrand in (3.13) take on their maximum values at $t_j=0$ and $t_j=1$.
The large-$N$ expansion can be obtained by expanding the integrand
about the maximum values.
Note in particular that for $t_j\sim 0$ and $L=2 \pi$
\begin{equation}
[1 - (1 - e^{2 \pi i \theta /L})t_j]^{N-2} \sim
e^{-(N-2)(1 - e^{i \theta})t_j}
\end{equation}
while for $t_j \sim 1$ and $L = 2 \pi$
\begin{equation}
[1 - (1 - e^{2 \pi i \theta /L})t_j]^{N-2} \sim
e^{i(N-2)\theta}e^{-(N-2)(1 - e^{-i \theta})(1 - t_j)}
\end{equation}

At this stage we observe that the problem is very similar to one we have
solved recently [12]. The solved problem is the task of calculating the
large-$x$ asymptotic expansion of the thermodynamic limit
form of the two-particle distribution. Now in the thermodynamic limit
$N, L \rightarrow \infty$ ($N/L = \eta$ fixed)  (3.13) becomes
$$
\rho(0,x)= \eta^2 (\eta x)^\beta A(\beta)
e^{- \pi i \beta \eta x}
\int_{[0,1]^\beta}dt_1 \dots dt_\beta \,
\prod_{j=1}^\beta e^{2 \pi i  t_j\eta x}
$$
\begin{equation}
\times
D_{2 /\beta,2 /\beta,4 /\beta}(t_1,\dots,t_\beta)
\end{equation}
from which the large-$x$ asymptotic expansion can be shown  to be of the form
$$
\rho^T(0,x)\sim \eta^2 \left [ -{1 \over \beta (\pi \eta x)^2}
+ {a^{(4)} \over (\eta x )^4} + \dots
+ \sum_{n=1}^{\beta/2} {\cos 2 \pi \eta x n \over (\eta x)^{4 n^2 /\beta}}
\left (b_n^{(0)} + {b_n^{(2)} \over (\eta x)^2} + \dots \right ) \right.
$$
\begin{equation}
+ \left. \sum_{n=1}^{\beta/2} {\sin 2 \pi \eta x n \over (\eta x)^{4 n^2 /\beta
+ 1}}
\left (c_n^{(0)} + {c_n^{(2)} \over (\eta x)^2} + \dots \right ) \right ]
\end{equation}
The method used in [12] to compute (3.18) was to
expand the integrand in (3.17) about the endpoints $t_j = 0,1$. We thus see
that
this problem is almost identical to the task of computing the large-$N$
expansion of (3.13).

Indeed, by comparing (3.13) and (3.17) and examining (3.15) and (3.16) we see
that
the large-$N$ expansion of (3.13) is identical to the large-$x$ expansion of
(3.17) and is thus given by (3.18) provided we make the replacements
\begin{equation}
\pi \eta x \mapsto 2 N \sin \theta /2, \qquad \eta \mapsto N/2 \pi.
\end{equation}
Substituting the resulting expansion in the l.h.s. of (3.2) we see that the
oscillatory terms all integrate to zero, and all but the first of
the non-oscillatory
terms vanish in the $N \rightarrow \infty$ limit. The first non-oscillatory
term gives the r.h.s. of (3.2) as required.

\vspace{1cm}
\noindent
{\bf Acknowledgement}

\noindent
I thank B. Jancovici for some useful remarks.

\pagebreak
\noindent
{\bf References}
\begin{description}
\item[][1] M.L. Mehta, {\it Random Matrices}, 2nd ed. (Academic Press,
San Diego, 1991).
\item[][2] E. Br\'ezin and A. Zee, Nucl. Phys. B 402 (1993) 613.
\item[][3] B. Eynard, hep-th/9401165, to appear Nucl. Phys. B.
\item[][4] C.W.J. Beenakker, preprint cond-mat/9310010, to appear Nucl. Phys.
B.
\item[][5] B.V. Bronk, J. Math. Phys. 6 (1965) 228.
\item[][6] T. Nagao and M. Wadati, J. Phys. Soc. Japan 61 (1992) 78, 1911;
T. Nagao and P.J. Forrester, preprint.
\item[][7] P.J. Forrester, Nucl. Phys. B 402 (1993) 709.
\item[][8] C.A. Tracy and H. Widom, Commun. Math. Phys. 161 (1994) 289.
\item[][9] C.W.J. Beenakker, Phys. Rev. Lett. 70 (1993) 1155.
\item[][10] A.M.S. Mac\^ede and J.T. Chalker, Phys. Rev. B 49 (1994) 4695.
\item[][11] B. Jancovici and P.J. Forrester, to appear Phys. Rev. B.
\item[][12] P.J. Forrester, Phys. Lett. A 179 (1993) 127.
\item[][13] W.K.H. Panofsky and M. Phillips, {\it Classical electricity and
magnetism}, 2nd ed. (Addison-Wesley, 1972).
\item[][14] P.J. Forrester, Nucl. Phys. B 416 (1994) 377.
\end{description}

\end{document}